\documentclass[12pt]{iopart}
\pdfoutput=1
\usepackage[utf8]{inputenc}
\usepackage[T1]{fontenc}

\expandafter\let\csname equation*\endcsname=\relax
\expandafter\let\csname endequation*\endcsname=\relax

\usepackage{amsmath, amssymb, color, graphicx, stmaryrd, wasysym}
\definecolor{linkcolor}{rgb}{0,0,0.6} 
\usepackage[colorlinks=true,
	pdfstartview = FitV,
	linkcolor    = linkcolor,
	citecolor    = linkcolor,
	urlcolor     = linkcolor,
	hyperindex   = true,
	hyperfigures = false]{hyperref}
\newcommand{\dd}{\text{d}}
\newcommand{\ee}{\text{e}}
\newcommand{\ii}{\text{i}}
\usepackage[]{todonotes}

\usepackage[numbers, square, sort&compress]{natbib}

\begin{document}

\title{Driven probe under harmonic confinement in a colloidal bath}

\author{Vincent D\'emery}
\address{Gulliver, CNRS, ESPCI Paris, PSL Research University, 10 Rue Vauquelin, 75005 Paris, France}
\address{Laboratoire de Physique, ENS de Lyon, Universit\'e Lyon, Universit\'e Claude Bernard Lyon 1, CNRS, F-69342 Lyon, France}

\author{\'Etienne Fodor}
\address{DAMTP, Centre for Mathematical Sciences, University of Cambridge, Wilberforce Road, Cambridge CB3 0WA, UK}

\begin{abstract}
Colloids held by optical or magnetic tweezers have been used to explore the local rheological properties of a complex medium and to extract work from fluctuations with some appropriate protocols. However, a general theoretical understanding of the interplay between the confinement and the interaction with the environment is still lacking. Here, we explore the statistical properties of the position of a probe confined in a harmonic trap moving at constant velocity and interacting with a bath of colloidal particles maintained at a different temperature. Interactions among particles are accounted for by a systematic perturbation, whose range of validity is tested against direct simulations of the full dynamics. Overall, our results provide a way to predict the effect of the driving and the environment on the probe, and can potentially be used to investigate the properties of colloidal heat engines with many-body interactions.
\end{abstract}

\maketitle


\newpage

\section{Introduction}

Optically trapped colloids are used to probe the rheology of complex media~\cite{Crocker2000, Meyer2006, Sriram2010, Ahmed2015, GomezSolano2015, Berner2018}, measure Casimir forces in critical mixtures~\cite{Paladugu2016, Martinez2017}, challenge the Landauer relation between information and thermodynamics~\cite{Berut2012} or test out-of-equilibrium extensions of the fluctuation-dissipation relation~\cite{GomezSolano2009, Mehl2010}. Recently, microscopic engines have been realized by varying cyclically the stiffness of the trap and the temperature~\cite{Blickle2012, Martinez2016b} or the activity~\cite{Krishnamurthy2016} of the surrounding bath. For this broad range of applications, it is desirable to have a general theoretical framework able to describe the statistical properties of a trapped particle in contact with a complex bath, including out-of-equilibrium situations.

In active microrheology~\cite{Wilson2011b}, a colloid ---the probe--- held by an optical trap is driven through the investigated medium, for instance a colloidal suspension~\cite{Meyer2006}; the position of the colloid in the trap allows one to determine the drag force felt by the colloid and to infer a rheological property of the medium. However, almost all the theoretical analysis of the motion of the probe assume either a constant velocity or a constant force driving mode~\cite{Squires2005, Khair2006, Demery2010, Demery2010a, Zia2010, Harrer2012, Swan2013, Demery2014, Puertas2014}. As these driving modes correspond to, respectively, the strong and weak trap limits~\cite{Swan2013}, a theory for an arbitrary trap stiffness would bridge the gap between former results.

For microscopic engines, it is crucial to quantify the effect of the bath on the trapped colloid to understand their behavior and optimize their efficiency. In a simple fluid, the colloid follows a Langevin equation in an external potential and the heat and work extracted from its motion can be calculated explicitly~\cite{Schmiedl2008}. In more complex fluids, such as a colloidal suspension or even a bacterial bath~\cite{Krishnamurthy2016}, explicit results are scarce~\cite{Kanazawa2014, Zakine2017, Martin2018}.

\begin{figure}
	\centering
	\includegraphics[width=.65\columnwidth]{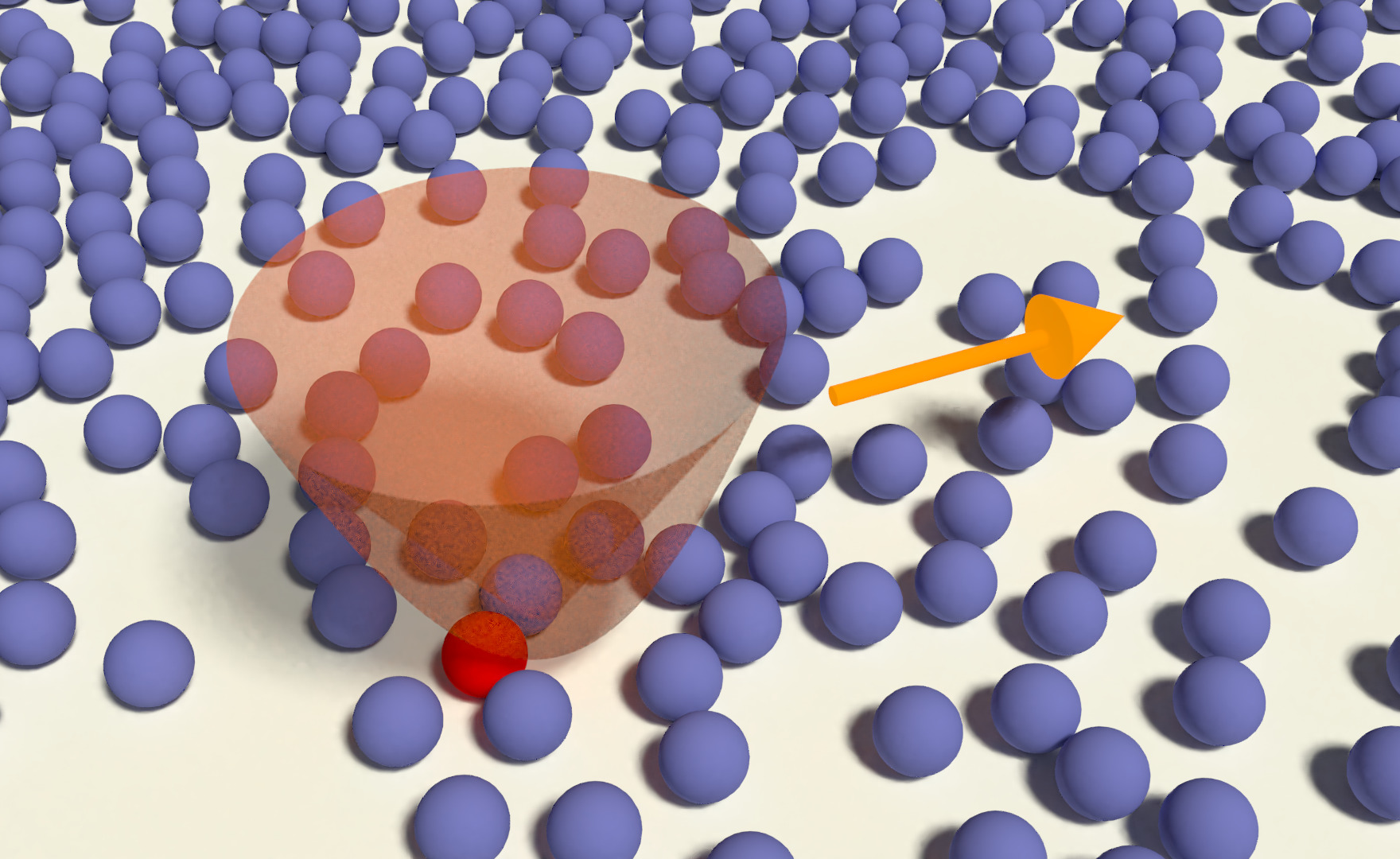}
	\caption{\label{fig:schema}
		Schematic representation of a probe (red) immersed in a bath of colloids (blue). The probe is confined in a harmonic potential whose centre position is displaced at constant velocity $\bf v$ with respect to the bath, as indicated by the arrow.
	}
\end{figure}

Here, we address the position statistics of a trapped probe with overdamped Langevin dynamics in contact with a general colloidal bath in two out-of-equilibrium situations~: (i) The trap is in motion with respect to the colloidal bath (Fig.~\ref{fig:schema}), which can be applied to active microrheology. (ii) The colloidal bath has a temperature different from the temperature felt by the probe, which can be used to model an active bath~\cite{Grosberg2015}. Treating the colloidal bath under the random phase approximation~\cite{Demery2014, Dean2014} and using a path-integral framework introduced in~\cite{Demery2011}, we compute perturbatively the effect of the bath on the generating function of the probe position. From the generating function, we can extract the average position, which is proportional to the drag force on the probe measured in active microrheology, and the variance of the position, which can be regarded as an effective temperature, and is related to force-induced diffusion in constant force microrheology~\cite{Zia2010}.

This article is organized as follows. The model used to describe the dynamics of the probe and its coupling to the bath is defined in Sec.~\ref{sec:model}. The calculation of the correction to the generating function of the position of the probe is computed in Sec.~\ref{sec:calc_gen_func}; the main result is the relation (\ref{eq:cumul_g}). The average and variance of the probe position are given in Sec.~\ref{sec:application} and applied to the situations (i) and (ii); limiting cases are discussed. Our results are compared to numerical simulations in Sec.~\ref{sec:num}.


\section{Model of a trapped probe in a colloidal bath}\label{sec:model}

We consider a probe particle with position ${\bf r}_0\in\mathbb{R}^d$ at temperature $T$, immersed in a bath of colloids with positions $\{{\bf r}_i\}_{i>0}$ at temperature $T_{\mathrm{b}}$ (Fig.~\ref{fig:schema}). The probe is confined in a harmonic trap with stiffness $\kappa$ located at the origin, corresponding to the potential $\kappa r^2/2$, and interact with the particles of the bath via the pairwise interaction potential $U(r)$; bath colloids interact via the pairwise interaction potential $V(r)$. In order to represent a motion of the trap with a velocity $\bf v$, the particles of the bath are advected with a velocity $-\bf v$. We assume that all the particles follow an overdamped Langevin dynamics, so that
\begin{align}
		\dot{\bf r}_0 &= - \mu \kappa {\bf r}_0 - \mu \nabla_0 \sum_i U ({\bf r}_i-{\bf r}_0) + \sqrt{2\mu T} {\boldsymbol\xi}_0 , \label{eq:dyn}
		\\
		\dot{\bf r}_i &= - {\bf v} - \nabla_i \Big[ \sum_j V({\bf r}_i-{\bf r}_j) + U({\bf r}_i-{\bf r}_0) \Big] + \sqrt{2T_{\mathrm{b}}} {\boldsymbol\xi}_i , \label{eq:dyn_bath}
\end{align}
where $\mu$ is the probe mobility, and we have set the mobility of the bath particles to unity. The term ${\boldsymbol\xi}_i$ is a zero-mean white Gaussian noise with correlations $\langle\xi_{i\alpha}(t)\xi_{j\beta}(0)\rangle = \delta_{ij}\delta_{\alpha\beta}\delta(t)$. 

To explore the interplay between interactions with the bath particles and harmonic confinement, our goal is to obtain the single-time statistics of the probe position ${\bf r}_0$ in the stationnary state, which is described by the cumulant generating function
\begin{equation}\label{eq:cumul}
	\psi_{\bf q} = \big\langle \ee^{ \ii{\bf q}\cdot{\bf r}_0 } \big\rangle .
\end{equation}
Without probe-bath interaction ($U=0$), the probe position follows a Gaussian distribution with zero mean and variance $\langle {\bf r}_0^2 \rangle = dT/\kappa$, and $\psi_{\bf q} = \ee^{-\kappa q^2/(2T)}$.


\section{Calculation of the probe position generating function}
\label{sec:calc_gen_func}

To compute $\psi_{\bf q}$, we follow the approach of~\cite{Demery2014}.
We linearize the dynamics of the bath using the Random Phase Approximation in (Sec.~\ref{sub:lin_dyn_bath}); then, we incorporate it in the probe equation of motion to obtain an effective dynamics (Sec.~\ref{sub:eff_dyn_probe}); next, we use a path-integral formulation of the dynamics of the probe to calculate observables perturbatively in the probe-bath interaction $U$ (Sec.~\ref{sub:path_int}); finally, we obtain the correction to the generating function $\psi_{\bf q}$ (Sec.~\ref{sub:gen_fun}), which is our main result~\eqref{eq:cumul_g}).


\subsection{Linearized dynamics of the bath}
\label{sub:lin_dyn_bath}

The bath colloids can be represented by their density field $\rho({\bf r},t) = \sum_{i\neq0} \delta[{\bf r}-{\bf r}_i(t)]$, whose dynamics is given by the Dean equation~\cite{Dean1996}:
\begin{equation}\label{eq:rho}
	\begin{aligned}
		(\partial_t-{\bf v}\cdot\nabla) \rho ({\bf r},t) &= \nabla \cdot \bigg[ \rho({\bf r},t) \nabla \bigg( \int  V({\bf r}-{\bf r}')\rho({\bf r}',t) \dd{\bf r}' + U[{\bf r}-{\bf r}_0(t)] \bigg) + T_{\mathrm{b}} \nabla \rho({\bf r},t) \bigg]
		\\
		&\quad + \nabla \cdot \bigg[ \sqrt{2 T_{\mathrm{b}}\rho({\bf r},t) } {\boldsymbol\Lambda}({\bf r},t) \bigg] ,
	\end{aligned}
\end{equation}
where $\boldsymbol\Lambda$ is a zero-mean Gaussian noise with correlations $\langle\Lambda_\alpha({\bf r},t)\Lambda_\beta({\bf r}',t')\rangle = \delta_{\alpha\beta}\delta({\bf r}-{\bf r}')\delta(t-t')$.

Following~\cite{Demery2014}, we linearize~\eqref{eq:rho} around the average value $\rho_0$, which corresponds to the Random Phase Approximation~\cite{Hansen2006} in the theory of liquids. The modes of density fluctuations $\phi_{\bf k}$, defined as $\phi_{\bf k}(t) = \int \ee^{-\ii{\bf k}\cdot{\bf r}} \big[ \rho({\bf r},t) - \rho_0 \big] \dd{\bf r}$, follow
\begin{equation}\label{eq:phi}
	\partial_t\phi_{\bf k} = - [ {\bf k}^2(T_{\mathrm{b}}+\rho_0V_{\bf k}) -\ii{\bf v}\cdot{\bf k}] \phi_{\bf k} - \rho_0 {\bf k}^2 U_{\bf k} \ee^{-\ii{\bf k}\cdot{\bf r}_0} + \sqrt{2\rho_0T_{\mathrm{b}}} \ii{\bf k}\cdot{\boldsymbol\Lambda}_{\bf k},
\end{equation}
where here and in what follows the subscript ${\bf k}$ refers to the Fourier mode of a given field. It can be integrated exactly as
\begin{align}
\phi_{\bf k}(t) &= \int {\cal G}_{\bf k}(t-t') \bigg[ \sqrt{2\rho_0T_{\mathrm{b}}} \ii{\bf k}\cdot{\boldsymbol\Lambda}_{\bf k}(t') - \rho_0 {\bf k}^2 U_{\bf k} \ee^{-\ii{\bf k}\cdot{\bf r}_0(t')} \bigg] \dd t' ,\\
{\cal G}_{\bf k}(t)& = \exp\big\{ - [{\bf k}^2(T_{\mathrm{b}}+\rho_0V_{\bf k})- \ii{\bf v}\cdot{\bf k} ] t\big\} \Theta(t).
\end{align}
where $\Theta$ is the Heaviside step function. The bath linearization should only be valid for weak interactions {\it a priori}, yet it has been shown that it can also provide some useful insight even beyond such a regime~\cite{Martin2018, Tociu2018}.


\subsection{Effective dynamics for the probe}
\label{sub:eff_dyn_probe}

The force exerted by the bath on the probe can then be written as
\begin{equation}
	\begin{aligned}
		-  \nabla_0 \sum_i U[{\bf r}_i(t)-{\bf r}_0(t)] &= - \int_{\bf k} \ii {\bf k} U_{\bf k} \ee^{\ii{\bf k}\cdot{\bf r}_0(t)} \phi_{\bf k}(t)
		\\
		&= \rho_0  \int_{\bf k} \ii {\bf k} |{\bf k}|^2 U_{\bf k}^2 \int {\cal G}_{\bf k}(t-t') \ee^{\ii{\bf k}\cdot[{\bf r}_0(t)-{\bf r}_0(t')]} \dd t' + {\boldsymbol\Gamma}[{\bf r}_0(t),t] ,
	\end{aligned}
\end{equation}
where $\int_{\bf k} = \int \dd{\bf k}/(2\pi)^d$. 
The first term embodies the effect of the probe on the bath, which resists the probe displacement: this a damping term in the probe dynamics. The second term $\boldsymbol\Gamma$ is a zero-mean Gaussian noise, which reflects the effect of the bath noise $\boldsymbol\Lambda$ in the probe dynamics; its correlations are
\begin{equation}
	\langle\Gamma_\alpha({\bf r},t)\Gamma_\beta({\bf r}',t')\rangle = \rho_0  T_{\mathrm{b}} \int_{\bf k} \frac{k_\alpha k_\beta U_{\bf k}^2}{T_{\mathrm{b}}+\rho_0V_{\bf k}} \ee^{ \ii{\bf k}\cdot({\bf r}-{\bf r}') - {\bf k}^2(T_{\mathrm{b}}+\rho_0V_{\bf k})  |t-t'|  + \ii{\bf v}\cdot{\bf k}(t-t')} .
\end{equation}
As a result, we have obtained an effective dynamics for the probe position ${\bf r}_0$ by integrating the bath degrees of freedom $\{{\bf r}_i\}$. This introduces memory effects in the probe dynamics, with an explicit dependence on the microscopic details of the surrounding bath particles. 
At variance with some previous works~\cite{Steffenoni2016, Maes2017, England2018}, our approach does not assume a slow relaxation of the tracer dynamics compared to the bath.

Our approach relies on the fact that the probe is linearly coupled to the density fluctuations $\phi_{\bf k}$, which are Gaussian fields with first order dynamics.
It can thus be generalized to the coupling to any field satisfying these properties; a general model and the corresponding results are given in~\ref{app:gen}.


\subsection{Path-integral formulation}
\label{sub:path_int}

The dynamic action $\cal A$ which rules the path probability of probe trajectories ${\cal P} \sim \ee^{-\cal A}$ can then be separated into (i) a free-motion contribution ${\cal A}_0$, and (ii) a contribution from interactions ${\cal A}_{\rm int}$. Using some standard path integral techniques~\cite{Martin1973, Dominicis1975, Demery2011}, they can be written in terms of the probe position ${\bf r}_0$ and the conjugated process $\bar{\bf r}_0$ as
\begin{equation}\label{eq:action}
	\begin{aligned}
		{\cal A}_0 &= \int \bar{\bf r}_0(t) \cdot \big[ \ii \dot{\bf r}_0(t) + \mu T \bar{\bf r}_0(t) \big] \dd t ,
		\\
		{\cal A}_{\rm int} &= \rho_0\mu \int \dd t\dd t'\int_{\bf k} U_{\bf k}^2 \ee^{ \ii{\bf k}\cdot[{\bf r}_0(t)-{\bf r}_0(t')]} {\cal G}_{\bf k}(t-t') \big[{\bf k}\cdot\bar{\bf r}_0(t)\big] \bigg\{ \frac{\mu T_{\mathrm{b}} \big[{\bf k}\cdot\bar{\bf r}_0(t')\big]}{T_{\mathrm{b}}+\rho_0V_{\bf k}} - {\bf k}^2\bigg\} .
	\end{aligned}
\end{equation}
Note that the interaction action scales with the probe-bath interaction as ${\cal A}_{\rm int}\sim U^2$.

We can now compute the correction to the generation function of the position of the probe~\eqref{eq:cumul} perturbatively in the interaction action, which corresponds to the regime of weak interactions between the tracer and the bath particles. To the first order in ${\cal A}_{\rm int}$, we have
\begin{equation}\label{eq:cumul_corr}
\psi_{\bf q} = \big\langle \ee^{ \ii{\bf q}\cdot{\bf r}_0 } \big\rangle 
= \big\langle \ee^{ \ii{\bf q}\cdot{\bf r}_0 }\big\rangle_0 - \big\langle{\cal A}_{\rm int} \ee^{ \ii{\bf q}\cdot{\bf r}_0 }\big\rangle_0,
\end{equation}
where $\langle\cdot\rangle_0$ denotes an average with respect to ${\cal A}_0$, and we have used that $\langle {\cal A}_{\rm int}\rangle_0=0$~\cite{Demery2011}.


\subsection{Generating function}
\label{sub:gen_fun}

Given that the probe statistics drawn from ${\cal A}_0$ is Gaussian, evaluating the correction in~\eqref{eq:cumul_corr} amounts to computing Gaussian integrals. The leading order reads $\big\langle \ee^{ \ii{\bf q}\cdot{\bf r}_0(0) } \big\rangle_0 = \ee^{ - \kappa{\bf q}^2/2T }$. The first order correction to $\psi_{\bf q}$ requires to evaluate some correlations between the probe position ${\bf r}_0$ and the conjugated process $\bar{\bf r}_0$. We defer to~\ref{app:corr} the detailed derivation, yielding
\begin{equation}\label{eq:cumul_g}
	\begin{aligned}
		&\psi_{\bf q} - \ee^{-\kappa{\bf q}^2/(2T)}
		\\
		&\,=- \rho_0\mu \ee^{-\kappa{\bf q}^2/(2T)} \int_0^\infty \dd t\dd t' \int_{\bf k} ({\bf k}\cdot{\bf q}) U_{\bf k}^2 \ee^{-\mu\kappa t} \bigg\{ {\bf k}^2 + \frac{\mu T_{\mathrm{b}} \ee^{-\mu\kappa t'}}{T_{\mathrm{b}}+\rho_0V_{\bf k}} \Big[{\bf k}^2 + ({\bf k}\cdot{\bf q}) \ee^{-\mu\kappa t} \Big] \bigg\}
		\\
		&\,\quad\times \exp\bigg\{ -\big[{\bf k}^2(T_{\mathrm{b}}+\rho_0V_{\bf k})-\ii{\bf v}\cdot{\bf k}\big] t' - \frac{T}{\kappa} \Big[ 1 - \ee^{-\mu\kappa t'} \Big] \Big[{\bf k}^2 + ({\bf k}\cdot{\bf q}) \ee^{-\mu\kappa t} \Big] \bigg\}.
	\end{aligned}
\end{equation}
This is our main result. The generalization to arbitrary linear field dynamics is presented in~\ref{app:gen}. 
In the next section, we use this result to compute the mean and the variance of the position of the probe.


\section{Average and variance of the probe position and applications}
\label{sec:application}

The moments of the probe statistics can be directly obtained from the cumulant generating function. Specifically, the average position $\langle r_{0\alpha}\rangle$ and the position variance $\langle r_{0\alpha}r_{0\beta}\rangle-\langle r_{0\alpha}\rangle\langle r_{0\beta}\rangle$ can be obtained from the generating function $\psi_{\bf q}$ through
\begin{align}
\langle r_{0\alpha}\rangle &= -\ii \frac{\dd \psi_{\bf q}}{\dd q_\alpha} \bigg|_{|{\bf q}|=0} , \label{eq:gen_avg}\\
	\langle r_{0\alpha}r_{0\beta}\rangle &= - 2 \frac{\dd^2 \psi_{\bf q}}{\dd q_\alpha \dd q_\beta} \bigg|_{|{\bf q}|=0}. \label{eq:gen_var}
\end{align}
In what follows, we discuss various special cases for the average and variance, and show that simple expressions can be obtained in the limits of strong and weak confinement.


\subsection{Average position and drag force}

From the explicit expression of $\psi_{\bf q}$, given at first order in~\eqref{eq:cumul_g}, and the relation~\eqref{eq:gen_var}, we obtain the average probe position in the trap:
\begin{equation}\label{eq:average}
	\begin{aligned}
		\langle r_{0\alpha}\rangle &= \frac{\ii\rho_0}{\kappa} \int_0^\infty \dd t \int_{\bf k} k_\alpha {\bf k}^2 U_{\bf k}^2 \bigg[ 1 + \frac{\mu T_{\mathrm{b}} \ee^{-\mu\kappa t}}{T_{\mathrm{b}}+\rho_0V_{\bf k}} \bigg]
		\\
		&\quad\times \exp\bigg\{ -\big[{\bf k}^2(T_{\mathrm{b}}+\rho_0V_{\bf k})-\ii{\bf v}\cdot{\bf k}\big] t - \frac{{\bf k}^2T}{\kappa} \Big[ 1 - \ee^{-\mu\kappa t} \Big] \bigg\}  .
	\end{aligned}
\end{equation}
It is non-zero only along the direction of the trap velocity ${\bf v}=v\hat{\bf e}$ as expected. The angular integral can be performed, so that the component parallel to $\hat{\bf e}$, $\langle r_{0\parallel} \rangle$, is
\begin{equation}\label{eq:average_int}
	\begin{aligned}
		\langle r_{0\parallel} \rangle &= -\frac{\rho_0}{\kappa} \int_0^\infty \dd t \int_0^\infty {\rm d}k {\cal K}_d(k,kvt) U_{\bf k}^2 \bigg[ 1 + \frac{\mu T_{\mathrm{b}} \ee^{-\mu\kappa t}}{T_{\mathrm{b}}+\rho_0V_{\bf k}} \bigg]
		\\
		&\quad \times\exp\bigg\{ -{\bf k}^2(T_{\mathrm{b}}+\rho_0V_{\bf k}) t - \frac{{\bf k}^2T}{\kappa} \Big[ 1 - \ee^{-\mu\kappa t} \Big] \bigg\},
	\end{aligned}
\end{equation}
where the explicit expression of ${\cal K}_d$ depends on the spatial dimension $d$:
\begin{equation}
	{\cal K}_d(k, u)=
	\begin{cases}
		\frac{k^3}{\pi} \sin(u) & \text{for $d=1$,}
		\\
		\frac{k^4}{2\pi} J_1(u) & \text{for $d=2$,}
		\\
		\frac{k^5}{2(\pi u)^2}\big[\sin(u) - u\cos(u)\big] & \text{for $d=3$,}
	\end{cases}
\end{equation}
$J_n$ referring to the Bessel function of order $n$.

We now address the weak and strong confinement regimes. The relative strength of the confinement is determined by the ratio of the probe and bath relaxation time scales: (i) the probe relaxes in the trap with a typical time $\tau_{\rm trap}\sim 1/(\mu\kappa)$, and (ii) the relaxation of the surrounding bath particles is controlled by the diffusive time $\tau_{\rm bath}\sim a^2/(T_{\mathrm{b}}+\rho_0 V_{a^{-1}})$, where $a$ is the characteristic size of the probe-bath interaction. The probe is weakly confined when $\tau_{\rm trap}\gg\tau_{\rm bath}$, or $\kappa\ll\kappa^* =(T_{\mathrm{b}}+\rho_0 V_{a^{-1}})/(\mu a^2)$, and strongly confined in the opposite limit.

In these asymptotic regimes, the average position gets simplified as
\begin{equation}\label{eq:avg_as}
	\begin{aligned}
		\langle r_{0\parallel} \rangle &\underset{\kappa \gg \kappa^*}{\simeq} -\frac{\rho_0 v }{\kappa} \int_{\bf k} \frac{k_\parallel^2 {\bf k}^2U_{\bf k}^2}{{\bf k}^4(T_{\mathrm{b}}+\rho_0V_{\bf k})^2 + ({\bf v}\cdot{\bf k})^2} ,
	\\
	\langle r_{0\parallel} \rangle &\underset{\kappa \ll \kappa^*}{\simeq} -\frac{\rho_0 v }{\kappa} \int_{\bf k} \frac{k_\parallel^2 {\bf k}^2U_{\bf k}^2}{{\bf k}^4(\mu T + T_{\mathrm{b}}+\rho_0V_{\bf k})^2 + ({\bf v}\cdot{\bf k})^2} \cdot \frac{(1+\mu)T_{\mathrm{b}}+\rho_0V_{\bf k}}{T_{\mathrm{b}}+\rho_0V_{\bf k}} .
	\end{aligned}
\end{equation}
Neither the probe temperature $T$ nor the probe mobility $\mu$ affect the average position for a strong confinement ($\kappa\gg\kappa^*$), showing that the probe position is essentially controlled by interactions with surrounding particles in this regime.

The asymptotic results~\eqref{eq:avg_as} can be related to previous studies on active microrheology, either at constant velocity~\cite{Demery2010} or constant force~\cite{Demery2014}, where the environment surrounding the tracer is also described by a Gaussian field. In steady state, the average drag force ${\bf f}_{\rm drag} = -\langle\nabla_0 \sum_i U({\bf r}_i-{\bf r}_0)\rangle$, which is exerted by surrounding particles on the probe, compensates the restoring force of the harmonic trap, so that ${\bf f}_{\rm drag} = \kappa \langle{\bf r}_0\rangle$. Substituting~\eqref{eq:avg_as} for a strong confinement and equal temperatures ($\kappa\gg\kappa^*$ and $T=T_{\mathrm{b}}$), one recovers the expression obtained previously in~\cite{Demery2010} for $\mu=1$. Thus, the strong confinement regime corresponds to the constant velocity driving mode.

The effective mobility of the probe can be defined as the ratio of its velocity to the total drag force, which includes the drag force from the colloidal bath, ${\bf f}_{\rm drag}$, and the drag due to the solvent, ${\bf f}_{\rm sol} = -\mu^{-1}{\bf v}$:
\begin{equation}
\mu_{\rm eff} = \frac{v}{|-\mu^{-1}v + f_{{\rm drag },\parallel}|} = \mu \left(1+\frac{\mu f_{{\rm drag},\parallel}}{v}\right)
\end{equation}
at first order in ${\bf f}_{\rm drag} \sim U^2$. Substituting~\eqref{eq:avg_as} for a weak confinement and equal temperatures ($\kappa\ll\kappa^*$ and $T=T_{\mathrm{b}}$), we recover the result derived in~\cite{Demery2014} for $\mu=1$, showing that the weak trap limit corresponds to the constant force driving mode.

\begin{figure}
	\centering
	\includegraphics[width=.7\columnwidth]{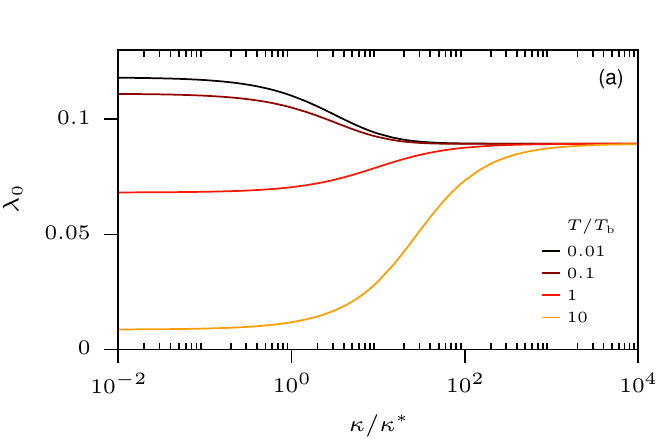}
	\vskip.5cm
	\includegraphics[width=.7\columnwidth]{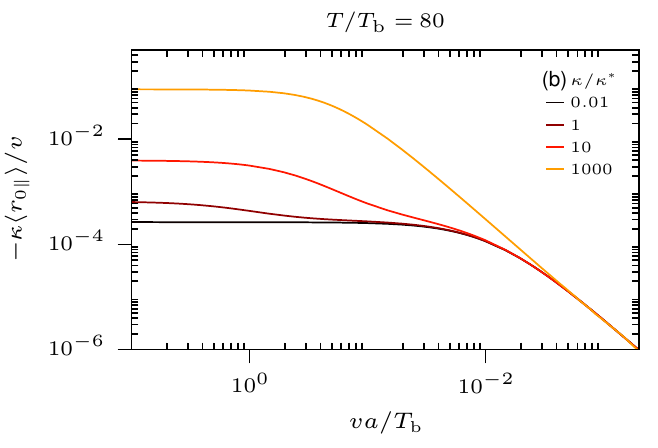}
	\caption{\label{fig:avg}
		(a)~Drag coefficient $\lambda_0 = \underset{v\to0}{\lim} -\kappa\langle r_{0\parallel}\rangle/v$ as a function of the scaled stiffness $\kappa/\kappa^*$ for different temperature ratios $T/T_{\rm b}$.
		(b)~Scaled tracer displacement $-\kappa\langle r_{0\parallel}\rangle/v$ as a function of the scaled trap velocity $v a/ T_{\rm b}$ for different scaled stiffnesses $\kappa/\kappa^*$.
		 The displacement $\langle r_{0\parallel}\rangle$ is given by evaluating numerically~\eqref{eq:average_int}, where the interaction potentials are both taken as a Gaussian core: $V({\bf r})=\varepsilon {\rm e}^{-|{\bf r}|^2/2a}$ and $U=V$.	Parameters: $\rho_0=0.1$, $\varepsilon=1$, $T_{\mathrm{b}}=1$, $\mu=1$, $a=1$, $d=2$.
	}
\end{figure}

The evolution of the drag coefficient $\lambda_0 = \underset{v\to0}{\lim} -\kappa\langle r_{0\parallel}\rangle/v$ as a function of trap stiffness shows a smooth crossover between the two asymptotic regimes, as reported in Fig.~\ref{fig:avg}(a). In the strong confinement regime ($\kappa\gg\kappa^*$), the drag is independent of the tracer temperature $T$, since the constant velocity driving mode is not affected by how the tracer interacts with the solvent. In contrast, the drag at weak confinement ($\kappa\ll\kappa^*$) decreases monotonically with $T/T_{\rm b}$. In this regime, increasing the tracer fluctuations stemming from the solvent, as controlled by $T$, effectively reduces the relative strength of interactions with bath particles, thereby decreasing their resistance to tracer motion. Moreover, when $T>T_{\rm b}$, we observe that the scaled displacement $-\kappa\langle r_{0\parallel}\rangle/v$ can exhibit a transient plateau in between the one at small velocity, where the drag coefficient is commonly defined, and the shear thinning regime at large velocity; see the curve for $\kappa=\kappa^*$ in Fig.~\ref{fig:avg}(b). This suggests that, for a regime of parameters, this intermediate plateau can potentially lead to underestimating the drag coefficient $\lambda_0$, namely if one does not evaluate $\lambda_0$ for sufficiently small velocity values.

Overall, our results draw a direct analogy between asymptotic confinement regimes and well established microrheology experiments. In particular, this suggests that confining a probe in a moving harmonic trap with tunable stiffness provides a way to quantify the crossover between standard microrheology methods, either at constant velocity or at constant force.


\subsection{Position variance and effective temperature}

The correction from the equilibrium probe variance follows from~\eqref{eq:cumul_g} and~\eqref{eq:gen_var} as
\begin{equation}
	\langle r_{0\alpha}r_{0\beta}\rangle-\langle r_{0\alpha}\rangle\langle r_{0\beta}\rangle = \delta_{\alpha\beta} \frac{T_\alpha^{\rm (eff)}}{\kappa} ,
\end{equation}
where we have introduced the effective temperature $T_\alpha^{\rm (eff)}$ defined as
\begin{equation}\label{eq:variance}
	\begin{aligned}
		\frac{T_\alpha^{\rm (eff)}}{T} &= 1 + \rho_0 \int_0^\infty \dd t \int_{\bf k} k_\alpha^2 U_{\bf k}^2 \bigg\{ \frac{\mu \ee^{-\mu\kappa t}}{T_{\mathrm{b}}+\rho_0V_{\bf k}}\cdot\frac{T_{\mathrm{b}}}{T} - \frac{{\bf k}^2}{\kappa} \Big[ 1 + \frac{\mu T_{\mathrm{b}} \ee^{-\mu\kappa t}}{T_{\mathrm{b}}+\rho_0V_{\bf k}} \Big]\Big[ 1 - \ee^{-\mu\kappa t} \Big] \bigg\}
		\\
		&\quad\times \exp\bigg\{ -\big[{\bf k}^2(T_{\mathrm{b}}+\rho_0V_{\bf k})-\ii{\bf v}\cdot{\bf k}\big] t - \frac{{\bf k}^2T}{\kappa} \Big[ 1 - \ee^{-\mu\kappa t} \Big] \bigg\}.
	\end{aligned}
\end{equation}
The term $\langle r_{0\alpha}\rangle\langle r_{0\beta}\rangle$ does not contribute at order $U^2$, since $\langle r_{0\alpha} \rangle\sim U^2$, so that the variance coincides with $\langle r_{0\alpha}r_{0\beta}\rangle$ at first order. The integration over time in~\eqref{eq:variance} cannot be done explicitly, and the effective temperature depends on the orientation in general: parallel, $T^{\rm (eff)}_\parallel$, or orthogonal $T^{\rm (eff)}_\perp$ to the velocity ${\bf v}$. Note that the term {\it effective temperature} is introduced here for convenience, yet it should not be regarded as a proper thermodynamic temperature. In particular, others have used a similar term for the ratio of spontaneous fluctuations to response function~\cite{Cugliandolo2011}, or as a frequency-dependent measure of the distance from equilibrium~\cite{Mizuno2007, Visco2015, Turlier2016, Ahmed2018}.

In a static trap ($v=0$), the effective temperature~\eqref{eq:variance} is isotropic, namely $T^{\rm (eff)}_\parallel = T^{\rm (eff)}_\perp = T^{\rm (eff)}$, and it can be written as
\begin{equation}
	\begin{aligned}\label{eq:variance_static}
		\frac{T^{\rm (eff)}}{T} = 1 + &\frac{\rho_0}{d}\bigg[\frac{T_{\mathrm{b}}}{T}-1\bigg] \int_0^\infty \dd t \int_{\bf k} {\bf k}^4 U_{\bf k}^2 \Big[ 1 - \ee^{-\mu\kappa t} \Big]
		\\
		&\quad\times \exp\bigg\{ -{\bf k}^2(T_{\mathrm{b}}+\rho_0V_{\bf k}) t - \frac{{\bf k}^2T}{\kappa} \Big[ 1 - \ee^{-\mu\kappa t} \Big] \bigg\}.
	\end{aligned}
\end{equation}
It is directly proportional to the difference between the probe temperature $T$ and the bath temperature $T_{\mathrm{b}}$, and it vanishes at equilibrium ($T=T_{\mathrm{b}}$), as expected from the equipartition theorem.

For a moving trap with equal temperatures ($v\neq0$ and $T=T_{\mathrm{b}}$), we expand the effective temperature at small velocity as
\begin{equation}\label{eq:var_exp}
	\begin{aligned}
		\frac{T^{\rm (eff)}_\parallel}{T} &= 1 + \Big(\frac{v}{v_0}\Big)^2 \int u_\parallel^4 \dd\Omega_d + {\cal O}\Big(\frac{v}{v_0}\Big)^4 ,
		\\
		\frac{T^{\rm (eff)}_\perp}{T} &= 1 + \Big(\frac{v}{v_0}\Big)^2 \int u_\parallel^2 u_2^2 \dd\Omega_d + {\cal O}\Big(\frac{v}{v_0}\Big)^4 ,
	\end{aligned}
\end{equation}
where $u_\parallel$ and $u_2$ are the components of the unit vector ${\bf u}$ along ${\bf v}$ and along a direction perpendicular to ${\bf v}$, and $\dd\Omega_d$ refers to the elementary solid angle in $d$ dimensions. 
The velocity factor $v_0$ is defined by
\begin{equation}
	\begin{aligned}
		\frac{1}{v_0^2} &= \frac{\rho_0}{2} \int_0^\infty \dd t \int_{\bf k} {\bf k}^4 U_{\bf k}^2 \bigg\{ \frac{\mu \ee^{-\mu\kappa t}}{T+\rho_0V_{\bf k}} - \frac{{\bf k}^2}{\kappa} \Big[ 1 + \frac{\mu T\ee^{-\mu\kappa t}}{T+\rho_0V_{\bf k}} \Big]\Big[ 1 - \ee^{-\mu\kappa t} \Big] \bigg\} .
		\\
		&\quad\times \exp\bigg\{ -{\bf k}^2(T+\rho_0V_{\bf k}) t - \frac{{\bf k}^2T}{\kappa} \Big[ 1 - \ee^{-\mu\kappa t} \Big] \bigg\} .
	\end{aligned}
\end{equation}
The angular integrals in~\eqref{eq:var_exp} can be computed in arbitrary dimension $d$, as shown in~\ref{app:angular}, yielding
\begin{equation}
	T^{\rm (eff)}_\parallel-T = 3 \big[T^{\rm (eff)}_\perp-T\big].
\end{equation}
As a result, the first correction to the anisotropic components of the probe variance are related independently of the microscopic details, namely for any interaction potentials $V$ and $U$, any trap stiffness $\kappa$, any mobility $\mu$ and any temperature $T$.
The increment of the probe position fluctuations as the trap is driven through the colloidal bath is reminiscent of the force-induced diffusion in constant force microrheology, and the anisotropy found here has been observed in other systems~\cite{Zia2010, Harrer2012, Benichou2013e}.

\begin{figure}
	\centering
	\includegraphics[width=.75\columnwidth]{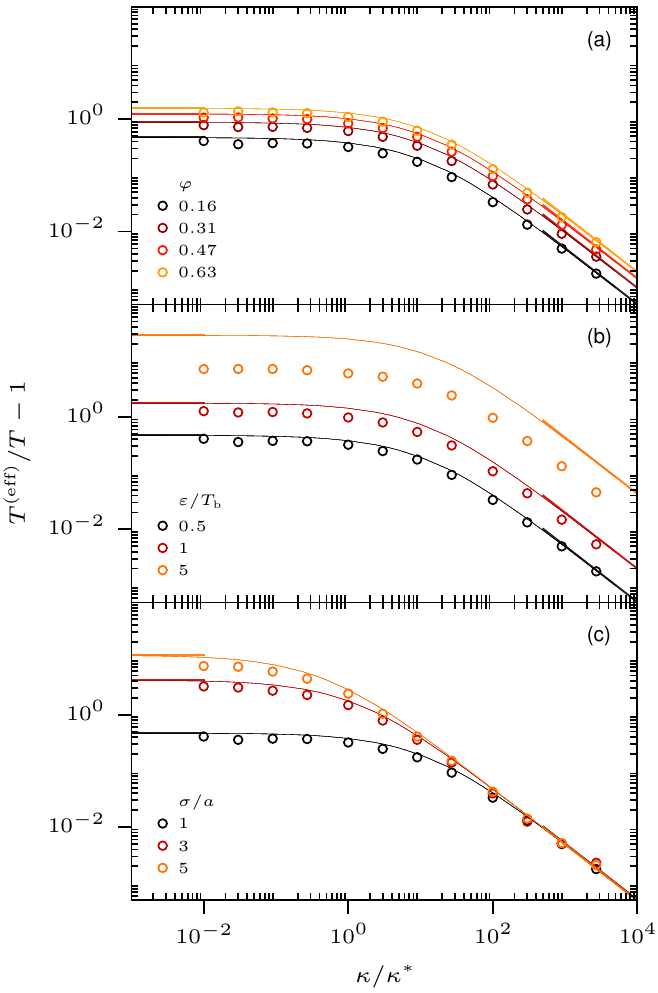}
	\caption{\label{fig:var}
		Correction to effective temperature $T^{\rm(eff)}/T-1$ as a function of the scaled stiffness $\kappa/\kappa^*$ for a static trap $v=0$. The circles are results from direct simulations of the microscopic dynamics~\eqref{eq:dyn}. The thin solid lines correspond to the numerical evaluation of our prediction~\eqref{eq:variance_static}, and the thick solid lines refer to the asymptotic behaviours in~\eqref{eq:var_as}.
		(a) Varying the packing fraction $\varphi$ at $\varepsilon=0.5T_{\mathrm{b}}$ and $\sigma=a$.
		(b) Varying the scaled interaction strength $\varepsilon/T_{\mathrm{b}}$ at $\varphi=0.16$ and $\sigma=a$.
		(c) Varying the scaled probe size $\sigma/a$ at $\varepsilon=0.5T_{\mathrm{b}}$ and $\varphi=0.16$.
		Other parameters: $T=10^{-3}$, $T_{\mathrm{b}}=1$, $\mu=1$.
	}
\end{figure}

Finally, the effective temperature can be simplified in the asymptotic confinement regimes as
\begin{equation}\label{eq:var_as}
	\begin{aligned}
		\frac{T^{\rm (eff)}_\alpha}{T} &\underset{\kappa\gg \kappa^*}{\simeq} 1 + \frac{\rho_0 T_{\mathrm{b}} }{\kappa T} \int_{\bf k} \frac{k_\alpha^2 U_{\bf k}^2}{T_{\mathrm{b}}+\rho_0V_{\bf k}} \cdot \frac{{\bf k}^2(1-T/T_{\mathrm{b}})(T_{\mathrm{b}}+\rho_0V_{\bf k})-\ii{\bf v}\cdot{\bf k}}{{\bf k}^2(T_{\mathrm{b}}+\rho_0V_{\bf k})-\ii{\bf v}\cdot{\bf k}},
		\\
		\frac{T^{\rm (eff)}_\alpha}{T} &\underset{\kappa \ll\kappa^*}{\simeq} 1 + \frac{\rho_0\mu T_{\mathrm{b}}}{T} \int_{\bf k} \frac{k_\alpha^2 U_{\bf k}^2}{T_{\mathrm{b}}+\rho_0V_{\bf k}} \cdot \frac{{\bf k}^2(1-T/T_{\mathrm{b}})(T_{\mathrm{b}}+\rho_0V_{\bf k})-\ii{\bf v}\cdot{\bf k}}{\big[{\bf k}^2(\mu T+T_{\mathrm{b}}+\rho_0V_{\bf k})-\ii{\bf v}\cdot{\bf k} \big]^2}.
	\end{aligned}
\end{equation}
The first order correction is negligible with respect to the leading order $T$ for a strong confinement ($\kappa\gg\kappa^*$), while the correction is of the same order in the weak confinement regime ($\kappa\ll\kappa^*$, Fig.~\ref{fig:var}).

\section{Numerical simulations}
\label{sec:num}

To probe the range of validity of these results, we measure the probe position variance from direct numerical simulations of the microscopic dynamics~\eqref{eq:dyn} for a static trap in a hot bath ($v=0$ and $T_{\mathrm{b}}\gg T$).
We consider particles in a two-dimensional box of size $L=40a$, with periodic boundary conditions, interacting through the soft-core potential $V({\bf r})=\varepsilon(1-|{\bf r}|/a)^2\Theta(a-|{\bf r}|)$, and we take identical interactions between probe and bath particles ($U=V$). To compare the simulation results with our predictions, the integration over the modes $\bf k$ is done with the corresponding Fourier potential $V_{\bf k} = (2\pi\varepsilon/{\bf k}^2)\big\{ \pi\big[J_1(a|{\bf k}|)H_0(a|{\bf k}|)-J_0(a|{\bf k}|)H_1(a|{\bf k}|)\big] - 2 J_2(a|{\bf k}|)\big\}$, where $J_n$ and $H_n$ respectively denote the Bessel and Struve functions of order $n$.

We extract the correction to the effective temperature $T^{\rm (eff)}/T-1$ at different values of the packing fraction $\varphi=\rho_0\pi a^4/4$. For weak interactions ($\varepsilon<T_{\mathrm{b}}$), our prediction~\eqref{eq:variance} indeed reproduces the measurements for a large range of trap stiffness values $\kappa$, up to the asymptotic regimes given in~\eqref{eq:var_as}, both at low and high packing fractions $\varphi$, as shown in Fig.~\ref{fig:var}(a). Some deviations become manifest when the strength of interactions $\varepsilon$ increases, as expected (Fig.~\ref{fig:var}(b)). Yet, we report the same qualitative behavior: the hot bath increases $T^{\rm (eff)}/T-1$  for a weak confinement $\kappa \ll\kappa^*$, and the correction to the equilibrium temperature $T$ is negligible for a strong confinement $\kappa \gg \kappa^*$, as shown in Fig.~\ref{fig:var}(b).

Finally, we consider simulations where the probe size $\sigma$, set by the range of the probe-bath interaction potential $U$, is larger than the size $a$ of bath particles. This leads to increase the variance correction~\eqref{eq:variance} for a weak confinement, yet the strong confinement regime is unaffected. Surprisingly, our perturbative treatment remains valid even for a correction to $T^{\rm (eff)}$ up to approximately ten times its equilibrium value $T$, as reported in Fig.~\ref{fig:var}(c). Overall, our numerical measurements support the validity of our perturbative approach, even in regimes where the correction to equilibrium is non-negligible.


\section{Conclusion}

We have studied the statistics of a Brownian probe immersed in a colloidal bath and confined in a harmonic trap for two out-of-equilibrium situations. From the perturbative calculation of the cumulant generating function of the probe position, we have deduced the average and variance, which are respectively related to the drag force and to an effective temperature. The validity of our approach, based on an explicit coarse-graining of the bath surrounding the tracer, is confirmed through direct simulations of the microscopic dynamics.

Some experimental realizations have demonstrated the feasibility of extracting work from the fluctuations of a colloidal probe~\cite{Blickle2011, Sood2016, Martinez2016, Martinez2017}. This is based on varying the parameters of an external confinement potential through cyclic protocols. For a quasistatic protocol with a harmonic trap, the work and the heat follow directly from the steady-state probe variance~\cite{Sekimoto1997, Seifert2012}. Therefore, our results could be used to explore the properties of heat engines operating in a colloidal bath. Besides, popular models of self-propelled particles generally involve a source of persistent fluctuations, whose correlations decay exponentially with time~\cite{Marchetti2013, Bechinger2016, Stark2016, Marchetti2018}. In our settings, the temperature difference between probe and bath can be regarded as the limiting case of a vanishing correlation time: this opens the door to considering engines operating with an active bath in such a regime.

Finally, exploring the finite-time properties of heat engines requires further knowledge of the probe dynamics, such as the two-time correlation of position~\cite{Sivak2012, Zulkowski2012, Zulkowski2015}. The maximal power has been computed recently when the relaxation in the trap is slower than the relaxation of the bath~\cite{Martin2018}, which corresponds to the weak trap limit. Regimes beyond this can now potentially be addressed with the tools developed here. More generally, the two-time correlation in a moving trap would also provide access to the viscoelastic properties of the fluid following some recent methods~\cite{GomezSolano2015, Berner2018}.


\section*{Acknowledgements}
	The authors acknowledge insightful discussions with Michael E. Cates. \'EF benefits from an Oppenheimer Research Fellowship from the University of Cambridge, and a Junior Research Fellowship from St Catherine's College.


\appendix

\section{Generic linear field dynamics}\label{app:gen}

We consider in this Appendix the case of a Brownian probe, with position ${\bf r}_0$, coupled with an arbitrary linear field $\phi$, whose dynamics is given by
\begin{equation}\label{eq:dyn_gen}
	\begin{aligned}
		\dot{\bf r}_0 &= -\mu\kappa{\bf r}_0 - \mu \int_{\bf k} {\rm i}{\bf k}K_{\bf k}{\rm e}^{{\rm i}{\bf k}\cdot{\bf r}_0} \phi_{\bf k} + \sqrt{2\mu T}{\boldsymbol\xi}_0 ,
		\\
		\partial_t\phi_{\bf k} &= - [R_{\bf k} A_{\bf k} - {\rm i}{\bf v}\cdot{\bf k}] \phi_{\bf k} - R_{\bf k}K_{\bf k}{\rm e}^{-{\rm i}{\bf k}\cdot{\bf r}_0} + \sqrt{2T_{\mathrm{b}}}{\boldsymbol\Xi}_{\bf k} ,
	\end{aligned}
\end{equation}
where we assume that $\{A_{\bf k}, R_{\bf k}, K_{\bf k}\}$ are symmetric with respect to ${\bf k}$, and $\boldsymbol\Xi$ is a zero-mean Gaussian noise with correlations	$\langle\Xi_\alpha({\bf r},t)\Xi_\beta({\bf r}',t')\rangle = \delta_{\alpha\beta}R({\bf r}-{\bf r}')\delta(t-t')$.
We recover the case of a colloidal bath in~\eqref{eq:phi} for $A_{\bf k}=T_{\mathrm{b}}/\rho_0+V_{\bf k}$, $R_{\bf k}=\rho_0{\bf k}^2$, and $K_{\bf k}=U_{\bf k}$. Following the procedure detailed in Sec.~\ref{sec:calc_gen_func}, the cumulant generating function of probe position $\psi_{\bf q}=\langle{\rm e}^{{\rm i}{\bf q}\cdot{\bf r}_0}\rangle$ can be obtained for the generic dynamics~\eqref{eq:dyn_gen} as
\begin{equation}
	\begin{aligned}
		\psi_{\bf q} &- \ee^{-\kappa{\bf q}^2/(2T)}
		\\
		&= - \mu  \ee^{-\kappa{\bf q}^2/(2T)} \int_0^\infty \dd t\dd t' \int_{\bf k} ({\bf k}\cdot{\bf q}) K_{\bf k}^2 \ee^{-\mu\kappa t} \bigg\{ R_{\bf k} + \frac{\mu T_{\mathrm{b}} \ee^{-\mu\kappa t'}}{A_{\bf k}} \Big[{\bf k}^2 + ({\bf k}\cdot{\bf q}) \ee^{-\mu\kappa t} \Big] \bigg\}
		\\
		&\quad\times \exp\bigg\{ -\big[R_{\bf k}A_{\bf k}-\ii{\bf v}\cdot{\bf k}\big] t' - \frac{T}{\kappa} \Big[ 1 - \ee^{-\mu\kappa t'} \Big] \Big[{\bf k}^2 + ({\bf k}\cdot{\bf q}) \ee^{-\mu\kappa t} \Big] \bigg\}.
	\end{aligned}
\end{equation}
This is the generalization of~\eqref{eq:cumul_g}. 
Note that this it corresponds to a perturbative calculation at order $K^2$.
The probe position average and variance follow as
\begin{equation}
	\begin{aligned}
		\langle r_{0\alpha}\rangle &= \frac{\ii}{\kappa} \int_0^\infty \dd t \int_{\bf k} k_\alpha K_{\bf k}^2 \bigg[ R_{\bf k} + \frac{\mu T_{\mathrm{b}}{\bf k}^2 \ee^{-\mu\kappa t}}{A_{\bf k}} \bigg]
		\\
		&\quad\times \exp\bigg\{ -\big[R_{\bf k}A_{\bf k}-\ii{\bf v}\cdot{\bf k}\big] t - \frac{{\bf k}^2T}{\kappa} \Big[ 1 - \ee^{-\mu\kappa t} \Big] \bigg\},
	\end{aligned}
\end{equation}
and
\begin{equation}
	\begin{aligned}
		\langle r_{0\alpha}&r_{0\beta}\rangle-\langle r_{0\alpha}\rangle\langle r_{0\beta}\rangle - \delta_{\alpha\beta} \frac{T}{\kappa}
		\\
		&= \frac{1}{\kappa} \int_0^\infty \dd t \int_{\bf k} k_\alpha k_\beta K_{\bf k}^2 \bigg\{ \frac{\mu T_{\mathrm{b}} \ee^{-\mu\kappa t}}{A_{\bf k}} - \frac{T}{\kappa} \Big[ R_{\bf k} + \frac{\mu T_{\mathrm{b}}{\bf k}^2 \ee^{-\mu\kappa t}}{A_{\bf k}} \Big]\Big[ 1 - \ee^{-\mu\kappa t} \Big] \bigg\} .
		\\
		&\quad\times \exp\bigg\{ -\big[R_{\bf k}A_{\bf k}-\ii{\bf v}\cdot{\bf k}\big] t - \frac{{\bf k}^2T}{\kappa} \Big[ 1 - \ee^{-\mu\kappa t} \Big] \bigg\}.
	\end{aligned}
\end{equation}
The expressions in the weak and strong confinement regimes, given in~\eqref{eq:avg_as} and~\eqref{eq:var_as} for the case of a colloidal bath, can be readily deduced.


\section{Cumulant generating function: perturbative treatment}\label{app:corr}

This Appendix is devoted to the derivation of the cumulant generating function of probe position $\psi_{\bf q}=\langle{\rm e}^{{\rm i}{\bf q}\cdot{\bf r}_0}\rangle$. From the explicit expression of the interacting action ${\cal A}_{\rm int}$, given in~\eqref{eq:action}, the first order correction $\big\langle{\cal A}_{\rm int}\ee^{\ii{\bf q}\cdot{\bf r}_0}\big\rangle_0$ in $\psi_{\bf q}$ requires to evaluate the following correlations
\begin{equation}
	\begin{aligned}
		\big\langle{\bf k}\cdot\bar{\bf r}_0(t)\ee^{\ii{\bf q}\cdot{\bf r}_0(0) + \ii{\bf k}\cdot[{\bf r}_0(t)-{\bf r}_0(t')]}\big\rangle_0 &= \big\langle\ii{\bf k}\cdot\bar{\bf r}_0(t) \big\{{\bf q}\cdot{\bf r}_0(0) + {\bf k}\cdot[{\bf r}_0(t)-{\bf r}_0(t')]\big\}\big\rangle_0
		\\
		&\quad\times \exp\bigg\{ -\frac{1}{2} \big\langle \big[ {\bf q}\cdot{\bf r}_0(0) + {\bf k}\cdot[{\bf r}_0(t)-{\bf r}_0(t')] \big]^2 \big\rangle_0 \bigg\} ,
		\\
		\big\langle\bar r_{0\alpha}(t)\bar r_{0\beta}(t')\ee^{\ii{\bf q}\cdot{\bf r}_0(0) + \ii{\bf k}\cdot[{\bf r}_0(t)-{\bf r}_0(t')]}\big\rangle_0 &= - \big\langle\bar r_{0\alpha}(t) \big\{{\bf q}\cdot{\bf r}_0(0) + {\bf k}\cdot[{\bf r}_0(t)-{\bf r}_0(t')]\big\}\big\rangle_0
		\\
		&\quad\times \big\langle\bar r_{0\beta}(t) \big\{{\bf q}\cdot{\bf r}_0(0) + {\bf k}\cdot[{\bf r}_0(t)-{\bf r}_0(t')]\big\}\big\rangle_0
		\\
		&\quad\times \exp\bigg\{ -\frac{1}{2} \big\langle \big[ {\bf q}\cdot{\bf r}_0(0) + {\bf k}\cdot[{\bf r}_0(t)-{\bf r}_0(t')] \big]^2 \big\rangle_0 \bigg\} ,
	\end{aligned}
\end{equation}
where we have applied Wick's theorem for some exponential observables~\cite{Demery2011, Demery2014}. The correlations of ${\bf r}_0$ and $\bar{\bf r}_0$ can be easily evaluated in the non-interacting dynamics as
\begin{equation}
	\begin{aligned}
		\big\langle r_{0\alpha}(t)r_{0\beta}(t')\big\rangle_0 &= \delta_{\alpha\beta} \frac{T}{\kappa} \ee^{-\mu\kappa|t-t'|} ,
		\\
		\big\langle r_{0\alpha}(t)\bar r_{0\beta}(t')\big\rangle_0 &= \ii\delta_{\alpha\beta} \ee^{-\mu\kappa(t-t')} \Theta(t-t') ,
		\\
		\big\langle [{\bf r}_0(t)-{\bf r}_0(t')]^2\big\rangle_0 &= \frac{2dT}{\kappa}\Big[ 1 - \ee^{-\mu\kappa|t-t'|} \Big] ,
	\end{aligned}
\end{equation}
yielding
\begin{equation}
	\begin{aligned}
		\big\langle{\bf k}\cdot\bar{\bf r}_0(t)\ee^{\ii{\bf q}\cdot{\bf r}_0(t) + \ii{\bf k}\cdot[{\bf r}_0(t)-{\bf r}_0(t')]}\big\rangle_0 &= - \Theta(-t)({\bf k}\cdot{\bf q}) \ee^{\kappa(\mu t-{\bf q}^2/2T)}
		\\
		&\quad\times \exp\bigg\{ - \frac{T}{\kappa} \Big[ 1 - \ee^{-\mu\kappa(t-t')} \Big] \Big[{\bf k}^2 + ({\bf k}\cdot{\bf q}) \ee^{\mu\kappa t} \Big] \bigg\} ,
	\end{aligned}
\end{equation}
and
\begin{equation}
	\begin{aligned}
		\big\langle\big[{\bf k}\cdot\bar{\bf r}_0(t)\big]\big[{\bf k}\cdot\bar{\bf r}_0(t')\big]&\ee^{\ii{\bf q}\cdot{\bf r}_0(t) + \ii{\bf k}\cdot[{\bf r}_0(t)-{\bf r}_0(t')]}\big\rangle_0
		\\
		&= \Theta(-t)({\bf k}\cdot{\bf q}) \ee^{\kappa(\mu t'-{\bf q}^2/2T)} \Big[{\bf k}^2 + ({\bf k}\cdot{\bf q}) \ee^{\mu\kappa t} \Big]
		\\
		&\quad\times \exp\bigg\{ - \frac{T}{\kappa} \Big[ 1 - \ee^{-\mu\kappa(t-t')} \Big] \Big[{\bf k}^2 + ({\bf k}\cdot{\bf q}) \ee^{\mu\kappa t} \Big] \bigg\} .
	\end{aligned}
\end{equation}
Substituting this in $\big\langle{\cal A}_{\rm int}\ee^{\ii{\bf q}\cdot{\bf r}_0(0)}\big\rangle_0$, we then deduce our final result for $\psi_{\bf q}$, written in~\eqref{eq:cumul_g} by using the change of variables $\{t,t-t'\}\to\{-t,t'\}$.


\section{Anisotropic probe variance: angular integrals}\label{app:angular}

In this Appendix, we compute explicitly the angular contribution of the anisotropic components of probe variance, when $v\neq0$ and $T=T_{\mathrm{b}}$, as given in~\eqref{eq:var_exp} for small $v$:
\begin{equation}
	{\cal I}_4 = \int u_\parallel^4 \dd\Omega_d ,
	\quad
	{\cal I}_{2,2} = \int u_\parallel^2 u_2^2 \dd\Omega_d .
\end{equation}
First, we note that ${\cal I}_4$ and ${\cal I}_{2,2}$ can be related in terms of ${\cal I}_2$, defined as
\begin{equation}
	{\cal I}_2 = \int u_\parallel^2 \dd\Omega_d = \int u_\parallel^2 \bigg[\sum_{i=1}^d u_i^2\bigg] \dd\Omega_d = {\cal I}_4 + (d-1){\cal I}_{2,2} ,
\end{equation}
where we have used that ${\bf u}$ is a unit vector. The integrals ${\cal I}_4$ and ${\cal I}_2$ can be deduced from a generic integral ${\cal I}_n$, given by
\begin{equation}
	{\cal I}_n = \int_0^\pi (\cos\theta)^n (\sin\theta)^{d-2} \dd\theta = B \bigg[ \frac{d-1}{2}, \frac{n+1}{2} \bigg] ,
\end{equation}
where we have introduced the Beta function $B(x,y)=\Gamma(x)\Gamma(y)/\Gamma(x+y)$, and $\Gamma$ is the Euler's Gamma function. The ratio ${\cal I}_{2,2}/{\cal I}_4$ follows as
\begin{equation}
	\frac{{\cal I}_{2,2}}{{\cal I}_4} = \frac{1}{d-1} \bigg[ \frac{{\cal I}_2}{{\cal I}_4}  - 1 \bigg] = \frac{1}{3} ,
\end{equation}
where we have used $\Gamma(d+1)=d\Gamma(d)$.



\bibliographystyle{iopart-num}
\bibliography{DrivenTracer_ref}

\end{document}